\documentclass[a4paper,12pt,english]{article}

\usepackage{amsfonts,bm,amssymb,array,babel,cite}
\usepackage[mathscr]{euscript}
\usepackage[all,color]{xy}
\usepackage{relsize}
\usepackage{tikz}
\usepackage{url}
\usepackage{amsmath,amsthm,bm}
\usepackage{accents}
\usepackage[]{graphicx}
\usepackage[T1]{fontenc}

\usepackage{empheq}
\usepackage{color}
\usepackage{mathtools}

\makeatletter

\newcommand{\dd}{\mathrm{d}}

\newcommand{\bbm}{\left(\begin{matrix}}
	\newcommand{\ebm}{\end{matrix}\right)}
\newcommand{\beq}{\begin{eqnarray}}
\newcommand{\eeq}{\end{eqnarray}}

\makeatother

\newcommand{\sfrac}[2]{{\textstyle\frac{#1}{#2}}}

\newcommand{\be}{\begin{equation}}
\newcommand{\ee}{\end{equation}}

\newcommand{\beqa}{\begin{eqnarray}}
\newcommand{\eeqa}{\end{eqnarray}} 
\def\nn{\nonumber} \def \bea{\begin{eqnarray}} \def\eea{\end{eqnarray}}

\newcommand{\barr}{\begin{array}}
	\newcommand{\earr}{\end{array}}
\numberwithin{equation}{section}
\newcommand{\mf}{\mathfrak}

 \def\g{\gamma} 
 \def\d{\delta} 

\def\e{\epsilon} 
    
\def\l{\lambda} \def\L{\Lambda}  \def\m{\mu}
\def\n{\nu}
\def\o{\omega}   
 \def\S{\Sigma}

\def\R{{\mathbb R}} \def\C{{\mathbb C}} \def\N{{\mathbb N}} 

 \def\one{\mbox{1 \kern-.59em {\rm l}}}

\def\bi{\begin{itemize}} \def\ei{\end{itemize}}

\def\n2{\bm\nabla^2}

\begin{document}
	
	\makeatother
	
	\parindent=0cm
	
	\renewcommand{\title}[1]{\vspace{10mm}\noindent{\Large{\bf
				
				#1}}\vspace{8mm}} \newcommand{\authors}[1]{\noindent{\large
			
			#1}\vspace{5mm}} \newcommand{\address}[1]{{\itshape #1\vspace{2mm}}}
	
	\begin{titlepage}
		
		
		\begin{center}
			
			\vskip 3mm

\title{{\large Noncommutative Gauge Theory and Gravity in Three Dimensions}}

\bigskip

\authors{Athanasios Chatzistavrakidis$^{\circ}$,\, Larisa Jonke$^{\circ}$,\, Danijel Jurman$^{\circ}$, \\ \vspace{5pt} George Manolakos$^{\bullet}$, \, Pantelis Manousselis$^{\bullet}$,\, George Zoupanos$^{\bullet,\star}$}

\bigskip

\address{$^{\circ}$ Theoretical Physics Division, Rudjer Bo$\check s$kovi\'c Institute, \\
	Bijeni$\check c$ka 54, 10000  Zagreb, Croatia
	
	\
	
	$^{\bullet}$ Physics Department, National Technical University, \\
	Zografou Campus, GR-15780 Athens, Greece

\

$^{\star}$ 	Max-Planck Institut f\"ur Physik, \\ F\"ohringer Ring 6, D-80805 Munich, Germany}
\end{center}

\vskip 2.5cm

\begin{center}
\textbf{Abstract}
\end{center}

The Einstein-Hilbert action in three dimensions and the transformation rules for the dreibein and spin connection can be naturally described in terms of gauge theory. In this spirit, we use
covariant coordinates in noncommutative gauge theory in order to describe 3D gravity in the framework of noncommutative geometry. We consider 3D noncommutative spaces based on SU(2) and SU(1,1), as  foliations of fuzzy 2-spheres and fuzzy 2-hyperboloids respectively. Then we
construct a U(2)$\times$ U(2) and a GL(2,$\C$) gauge theory on them, identifying the corresponding noncommutative vielbein and spin connection.  We determine the transformations of the fields and an action in terms of a matrix model and discuss its relation to 3D gravity.

\end{titlepage}


\section{Introduction}

A popular idea addressing the problem of our lack of knowledge for the quantum structure of spacetime is that at small distances its coordinates exhibit a noncommutative structure.
Once this idea is taken seriously, naturally one asks which are its implications for gravity.
A modest approach is to draw lessons from  general relativity and study the modifications or corrections to the symmetries, field equations and their solutions due to this noncommutativity of coordinates. 

It is interesting to recall one relation between gravity and gauge theories \cite{Utiyama:1956sy, Kibble:1961ba, MacDowell:1977jt, Kibble:1985sn}; general relativity, with or without cosmological constant, is obtained upon gauging the Poincar\'e or
(A)dS algebra. This is true for the transformation rules of the gauge fields---the vielbein and the spin connection---in any dimension. However, in three dimensions not only the transformation rules but also the dynamics, the Einstein-Hilbert action, can be described by gauge theory \cite{AchucarroTownsend,WittenCS}. Motivated by the existence of noncommutative gauge theories \cite{Madore:2000en},
one could then ask whether they can be used as a guide to noncommutative gravity. Such an approach was followed before, for example in Refs. \cite{Chamseddine:2000si,Chamseddine:2003we,Aschieri1,Aschieri2,Ciric:2016isg}. Similarly,
this has been studied also in three dimensions, utilizing the relation to Chern-Simons gauge theory
\cite{Cacciatori:2002gq,Cacciatori:2002ib,Aschieri3, Banados:2001xw}. The common feature of the
above works is that the noncommutative deformation is constant (Moyal-Weyl) and the analysis
is made using the corresponding $\star$-product and the Seiberg-Witten map \cite{Seiberg:1999vs}.

Alternatively, one can use another type of noncommutative geometries, matrix geometries, in order to explore quantum gravity \cite{Banks:1996vh,Ishibashi:1996xs}. Several approaches have been suggested in recent years, mainly based on Yang-Mills matrix models \cite{Yang:2006dk,Steinacker:2010rh,Kim:2011cr,Nishimura:2012xs,Furuta:2006kk,Hanada:2005vr,Aoki:1998vn,Nair:2006qg,Abe:2002in,Valtancoli:2003ve,Nair:2001kr}, pointing once more at direct
relations among noncommutative gauge theories and gravity. 
For another approach see Refs. \cite{Buric:2006di,Buric:2007zx,Buric:2007hb}, where a solid indication that the degrees of freedom or basic modes of the resulting theory of gravity can be put in correspondence with those of the noncommutative structure has been presented. In this case, the usual symmetries such as coordinate invariance are built-in, and the commutator of coordinates can have arbitrary dependence on them. 
 Moreover, describing gravity as a gauge theory in the context of matrix geometry is further motivated by gauge theories defined on fuzzy spaces \cite{Aschieri:2003vyAschieri:2004vhAschieri:2005wm}. Notably, the reduction of higher-dimensional gauge theories over fuzzy manifolds, used as extra dimensions, leads to renormalizable theories in four dimensions \cite{Aschieri:2006uwSteinacker:2007ay}, a feature that is worth exploring in the case of gravity too.

In an attempt to formulate gravity in the noncommutative setup, the price one has to pay is that noncommutative deformations generically break Lorentz invariance. For certain types of noncommutative spaces, it is possible to define deformed symmetries which are preserved, as for example in the case of $\kappa$-Minkowski spacetime \cite{Lukierski:1991pn,Lukierski:1992dt}, which appears as a solution of the Lorentzian IIB matrix model in Ref. \cite{Kim:2011ts}. However, there are special
 types of deformations, in fact some of the very first noncommutative geometries ever considered, that constitute covariant noncommutative spacetimes \cite{Yang:1947ud,Snyder:1946qz}.
 This spirit was recently revived in Ref. \cite{Heckman:2014xha}, where the authors discuss a realization of this idea and construct a noncommutative deformation of a general conformal field theory defined on four-dimensional dS or AdS spacetime. Similar four-dimensional constructions were pursued in Refs. \cite{Buric:2015wta,Sperling:2017dts,Steinacker:2016vgf,Buric:2017yes}.

 In this paper, we revisit 3D gravity from the point of view of noncommutative gauge theory. In order to do so, first we need to identify 3D noncommutative spaces with the appropriate symmetry. In the Euclidean case, an example of such a space is the foliation of 3D space with Euclidean signature by fuzzy spheres, first considered in Ref. \cite{Hammou:2001cc}. This space has the Lie algebra SU(2) as the algebra of noncommutative coordinates. However, unlike the fuzzy sphere \cite{Madore:1991bw, Hoppe}, one does not restrict the Hermitian matrices to be proportional to the generators of SU(2) in an irreducible representation, but instead one considers reducible representations and constructs large block-diagonal matrices with each block being an irreducible representation. It is in this sense that a third dimension opens up and each block of the matrix corresponds to a fuzzy sphere, each fuzzy sphere being a leaf of the foliation. This space has a natural action of SO(4), see e.g. \cite{Kovacik:2013yca}, which we would like to gauge. As usual in non-Abelian noncommutative gauge theories, additional generators have to be included in order to close the anticommutation relations, thus the gauge theory we consider in this case is U(2)$\times$ U(2) in a fixed representation.

 In the more interesting Lorentzian case, the analogous construction involves a 3D space with underlying symmetry
 SO(1,2). Such a space can be constructed in a similar fashion to the Euclidean case, as a foliation of 3D spacetime with Lorentzian signature by the 2D fuzzy hyperboloids
 of Ref. \cite{Jurman:2013ota}. On this space, one would like to consider a SO(3,1) gauge theory; for the same reason as before, the gauge group has to be extended to GL(2;$\C$), as for example in Ref. \cite{Aschieri1}, which gives the vielbein and the spin connection, as well as a pair of additional gauge fields.

 The paper is organized as follows. In Section 2, we briefly recall the construction of 
 3D gravity as a gauge theory of the Poincar\'e, de Sitter or Anti de Sitter algebra, depending on the cosmological constant. In section 3, we remind the reader of the standard construction of noncommutative gauge theories in terms of covariant coordinates, with special attention to the non-Abelian case. In Section 4, we discuss the 3D fuzzy spaces on which we will consider noncommutative gauge theories. These are based on the algebras of SU(2) and SU(1,1), respectively. We discuss in some detail the compact case and the finite-dimensional representations that define the fuzzy space as a discrete foliation of fuzzy 2-spheres. The non-compact case is argued to exist in a similar fashion, as a  foliation of 2D fuzzy hyperboloids. In Section 5, we discuss gauge theories on these spaces, in particular we present in detail a $\text{GL}(2;\C)$ gauge theory in the non-compact case, for which we derive the transformation rules of the noncommutative vielbein and spin connection, we construct the corresponding curvatures as well, and then we compare them with the commutative case. In section 6, we argue for a matrix model that yields the corresponding action, comparing it with the classical Chern-Simons action of 3D gravity. 
  Section 7 contains our conclusions.

\section{Gravity as a gauge theory}

\subsubsection*{3D gravity as a gauge theory}

The first order formulation of general relativity in three dimensions can be understood as a gauge theory of the Poincar\'e algebra $\mf{iso}(1,2)$.{\footnote{This is also true in any dimension for what concerns the transformation rules of the fields.}} In the presence of a cosmological constant,
the relevant algebras are the de Sitter and Anti de Sitter ones, $\mf{so}(1,3)$ and $\mf{so}(2,2)$
respectively.
The corresponding generators are the ones of local translations $P_a, a=1,2,3$ and the Lorentz transformations $J_{ab}$,
satisfying the commutation relations{\footnote{We employ the standard convention that
antisymmetrizations are taken with weight 1.}} (CRs)
\bea
[J_{ab},J_{cd}]=4\eta_{[a[c}J_{d]b]}~,\quad
{[}P_a,J_{bc}]=2\eta_{a[b}P_{c]}~,\quad
{[}P_a,P_b]=\Lambda J_{ab}~,\label{cr}
\eea
where $\eta_{ab}$ is the (mostly plus) Minkowski metric and $\Lambda$ denotes the cosmological constant.
These commutation relations are valid in any dimension, however a convenient rewriting in three dimensions
reads as
\be
[J_{ a},J_{ b}]= \epsilon_{ a b c}J^{ c}~,\quad
{[}P_{ a},J_{ b}]= \epsilon_{ a b c}P^{ c}~,\quad
{[}P_{ a},P_{ b}]=\L \e_{abc}J^c~,
\ee
using the definition $J^a=\sfrac 12 \epsilon^{abc}J_{bc}$.
 Gauging proceeds with the introduction of a gauge field for each algebra generator,
in particular the dreibein $e_{\mu}{}^a$ for translations and the spin connection $\omega_{\mu}{}^a=\sfrac 12\epsilon^{abc}\omega_{\m bc}$ for Lorentz transformations.
The gauge connection is collectively given as
\be
A_{\m}=e_{\m}{}^a(x)P_a+ \omega_{\mu}{}^{a}(x)J_{a}~,
\ee
and it transforms in the adjoint representation according to the standard rule
\be
\d A_{\m}=\partial_{\m}\epsilon+[A_{\m},\epsilon]~,
\ee
where the gauge transformation parameter is taken to be
\be
\epsilon=\xi^a(x) P_a+ \l^{a}(x)J_{a}~.
\ee
Thus one can find the transformations of the vielbein and spin connection,
\bea\label{deltae}
\d e_{\m}{}^a&=&\partial_{\m}\xi^a-\epsilon^{abc}\left(\xi_b\omega_{\m c}+\l_{b}e_{\m c}\right)~,
\\ \label{deltao}
\d \omega_{\m}{}^{a}&=&\partial_{\m}\l^{a}-\epsilon^{abc}\left(\l_b\omega_{\m c}+\Lambda \xi_b e_{\mu c}\right)~,
\eea
and their curvatures, using the usual formula
\be
R_{\m\nu}(A)=2\partial_{[\m}A_{\nu]}+[A_{\m},A_{\nu}]~.
\ee
Writing $R_{\m\nu}(A)=T_{\m\nu}{}^aP_a+  R_{\m\nu}{}^{a}J_{a}$\,, these turn out to be
\bea\label{tmn}
T_{\m\nu}{}^a&=&2\partial_{[\m}e_{\nu]}{}^a+2\epsilon^{abc}\omega_{[\mu b}e_{\nu]c}~,\\ \label{rmn}
R_{\m\nu}{}^{a}&=&2\partial_{[\m}\omega_{\nu]}{}^{a}+\epsilon^{abc}\left(\omega_{\m b}\omega_{\nu c}+\Lambda e_{\mu b}e_{\nu c}\right)~.
\eea
The Einstein-Hilbert action with or without cosmological constant $\Lambda$ in three dimensions,
\be \label{eh3}
S_{\text{EH3}}=\frac 1{16\pi G} \int_M  \epsilon^{\mu\nu\rho}\left(e_{ \mu}{}^{a}\left(\partial_{\nu}\omega_{\rho a}-\partial_{\rho}\omega_{\nu a}\right)+\epsilon_{a b c}e_{\mu}{}^a\o_{\nu}{}^b\o_{\rho}{}^c+\sfrac 13 \Lambda \e_{abc}e_{\mu}{}^ae_{\nu}{}^be_{\rho}{}^c
\right)\,,
\ee
is identical to the action functional of a Chern-Simons gauge theory of the Poincar\'e, dS or AdS algebra, upon choice of an appropriate
quadratic form in the algebra \cite{AchucarroTownsend,WittenCS}. The standard choice is
\be
\text{tr}(J^aP^b)=\delta^{ab}~, \quad \text{tr}(P^aP^b)=\text{tr}(J^aJ^b)=0~.
\ee
However, in three dimensions and for non-vanishing $\Lambda$, there exists a second non-degenerate invariant quadratic form, given as
\be \label{tr2}
\text{tr}(J^aP^b)=0~, \quad \sfrac 1{\Lambda}\text{tr}(P^aP^b)=\text{tr}(J^aJ^b)=\d^{ab}~,
\ee
yielding a different, yet classically equivalent, action \cite{WittenCS}. This second set of traces will be important in our study too.

\subsubsection*{Remarks on 4D gravity}

For completeness of the presentation, recall that the vielbein formulation of general relativity in four dimensions is associated to the Poincar\'e algebra $\mf{iso}(1,3)$ \cite{Utiyama:1956sy, Kibble:1961ba, MacDowell:1977jt, Kibble:1985sn}.
The latter comprises ten generators, the four generators of local translations $P_a, a=1,2,3,4,$ and the six Lorentz transformations $J_{ab}$,
satisfying the commutation relations \eqref{cr} for $\L=0$.  
Gauging proceeds in the same way, in terms of the vierbein $e_{\mu}{}^a$  and the spin connection $\omega_{\m}{}^{ab}$,
leading to the transformations
\bea
\d e_{\m}{}^a&=&\partial_{\m}\xi^a+\omega_{\m}{}^{ab}\xi_b-\l^a{}_be_{\m}{}^b~,
\\
\d \omega_{\m}{}^{ab}&=&\partial_{\m}\l^{ab}-2\l^{[a}{}_c\omega_{\m}{}^{\underline{c}b]}~,
\eea
for gauge parameters $(\xi^a,\l^{ab})$, and the curvatures
\bea
R_{\m\nu}{}^a(e)&=&2\partial_{[\m}e_{\nu]}{}^a-2\omega_{[\mu}{}^{ab}e_{\nu]b}~,\\
R_{\m\nu}{}^{ab}(\omega)&=&2\partial_{[\m}\omega_{\nu]}{}^{ab}-2\omega_{[\m}{}^{ac}\omega_{\nu]c}{}^b~.
\eea
Imposing vanishing torsion, $R_{\m\nu}{}^a(e)=0$, leads to the solution of the spin
connection in terms of the vielbein components.
The dynamics follows from the Einstein-Hilbert action
\be
S_{\text{EH4}}=\frac 12 \int \dd^4x\,\epsilon^{\mu\nu\rho\sigma}\epsilon_{abcd}\, e_{\mu}{}^{a}e_{\nu}{}^b R_{\rho\sigma}{}^{cd}(\omega)~.
\ee
However, in this case the action does not follow from gauge theory \cite{WittenCS}; thus only the kinematics of four-dimensional gravity is directly captured by gauge theory.

\section{Gauge theories on noncommutative spaces}
\label{sec3}

Gauge theories also exist on noncommutative spaces.
In that case, one begins with an algebra $\cal A$ of operators $X_{\mu}$, and refers to them as a noncommutative space with
noncommutative coordinates. The operators $X_{\mu}$ satisfy a commutation relation which is generically given as
\be
[X_{\m},X_{\nu}]=i\theta_{\m \nu}~.
\ee
Note that in general the dependence of $\theta_{\m \nu}$ is not a priori specified; there are two reasons for this. First, although customarily
it is assumed to depend on the noncommutative coordinates $X_{\m }$, it could in principle
as well depend on the corresponding momenta $P_{\m }$
(see Ref. \cite{Szabo:2009tn} for some examples). Second, one may wish to consider noncommutative spaces where the operators $X_{\m }$ do not close, as
for example the case of the fuzzy 4-sphere of Ref. \cite{Castelino:1997rv}, or more generally spaces where the tensor
$\theta_{\mu\nu}$ is
not fixed, as for example in Refs. \cite{Yang:1947ud,Heckman:2014xha}.

A natural way to introduce noncommutative gauge theories is through covariant noncommutative coordinates, which in this paper will be denoted as ${\cal X}_{\m }$ \cite{Madore:2000en}.
They are defined as
\be
{\cal X}_{\mu}=X_{\mu}+A_{\mu}~,
\ee
such that they obey a covariant gauge transformation rule,
\be\label{gentrafo}
\d {\cal X}_{\mu}=i[\epsilon,{\cal X}_{\mu}]~.
\ee
The quantity $A_{\mu}$ transforms as the analog of a gauge connection, 
\be\d A_{\mu}=-i[X_{\mu},\e ]+i[\e,A_{\mu}]~,\ee 
and it can be used to
define a noncommutative covariant field strength, which is in turn used to define the noncommutative gauge theory. The field strength is given in terms of covariant coordinates as
\be
F_{\mu\nu}=[{\cal X}_{\mu},{\cal X}_{\nu}]-i\bar{\theta}_{\mu\nu} \quad \mbox{and} \quad  F_{\mu\nu}=[{\cal X}_{\mu},{\cal X}_{\nu}]-iC_{\mu\nu\rho}{\cal X}_{\rho}~,\label{fmn}
\ee
in the popular cases of constant $\bar{\theta}_{\mu\nu}$ and linear (Lie-type) noncommutativity respectively. In case $\theta_{\mu\nu}$ is not fixed, such a field strength has to be determined accordingly.

Up to now it was not specified whether the gauge theory is Abelian or not. This depends on the nature of the gauge parameters $\e$; if they are simply elements of the algebra ${\cal A}$, then the theory is Abelian, while when they are valued in a matrix algebra $\text{Mat}(\cal A)$, the theory is non-Abelian \cite{Madore:2000en}.
Whenever a non-Abelian
gauge theory is considered, there is an issue with the algebra where the gauge fields are valued.
In particular, if for the moment we collectively denote the generators of the algebra as $T^A$ and suppress the $\mu$ index,  the well-known relation
\be \label{nonAbelianCR}
[\epsilon,A]=[\epsilon^AT^A,A^BT^B]=\sfrac 12 \{\epsilon^A,A^B\}[T^A,T^B]+\sfrac 12 [\epsilon^A,A^B]\{T^A,T^B\}~,
\ee
reveals that restricting to a matrix algebra is not generically possible \cite{Madore:2000en}. This is due to the fact that the very last term, although
trivially vanishing in the commutative case, it is neither zero nor a Lie algebra element in the noncommutative case. This
is evaded either by considering the universal enveloping algebra \cite{Jurco:2000ja}, or by appropriately extending the generators and/or fixing the representation
such that the anticommutators close. In the following we shall consider such non-Abelian gauge theories on noncommutative spaces and we shall employ the second option.

\section{Three-dimensional fuzzy spaces based on SU(2) and SU(1,1)}
\label{sec4}

Our purpose is to write a gauge theory in the spirit of Section 3, which is associated to gravity in 3D. In order to achieve this goal, we should first identify the corresponding fuzzy space. 
In the Euclidean case, we consider the three-dimensional noncommutative space $\mathbb{R}^{3}_{\lambda}$ 
introduced in \cite{Hammou:2001cc}, which can viewed as a direct sum of fuzzy spheres. 
The fuzzy sphere  \cite{Madore:1991bw, Hoppe} is defined in terms of three rescaled angular momenta $X_i=\l J_i$, the Lie algebra 
generators of a unitary irreducible representation $D^j$ of SU(2), which satisfy
\be
[X_i,X_j]=i\lambda \epsilon_{ijk}X_k~,\quad \sum_{i=1}^3 X_iX_i=\l^2 j(j+1):=r^2~,\;\lambda \in \mathbb{R},\;2j\in \mathbb{N}.
\ee 
They  can be explicitly written in the standard basis of $\text{Mat}_{2j+1} ({\C})\cong \text{End}(\mathbb{C}^{2j+1})$ as:
\begin{eqnarray}\label{gensu2a}
X_{1} &=& \frac{\lambda}{2} \sum_{m} \left(\sqrt{(j+m)(j-m+1)}e^{j}_{m, m-1} + \sqrt{(j-m)(j+m+1)}e^{j}_{m, m+1}\right)~,  \\\label{gensu2b}
X_{2} &=& \frac{\lambda}{2i} \sum_{m} \left(\sqrt{(j+m)(j-m+1)}e^{j}_{m, m-1} - \sqrt{(j-m)(j+m+1)}e^{j}_{m,m+1}\right)~, \\\label{gensu2c}
X_{3} &=& \lambda \sum_{m} m e^{j}_{mm}~,
\end{eqnarray}
where the elementary matrices $ \{ e^{j}_{mn} \} , -j\leq m,n \leq j $ with matrix elements $ (e^{j}_{mn})_{kl} = \delta_{mk} \delta_{nl} $, corresponding 
as endomorphisms to operators $|jm\rangle \langle jn|$,  
are orthonormal  with respect to the inner product  $(a,b) = \text{Tr}(a^{\dagger}b), \forall a,b \in \text{Mat}_{2j+1} ({\C})$.
However, in the context of the fuzzy sphere, it is useful to introduce yet another basis of $\text{Mat}_{2j+1} ({\C})$, 
which is built from matrix polynomials in the generators (\ref{gensu2a})-(\ref{gensu2c}). Namely, due to the irreducibility of the representation $D^j$,
there are exactly $(2j+1)^2$ linearly independent completely symmetric polynomials in $X_i$ and a convenient way to choose orthogonal 
combinations denoted as $\mathbb{Y}^J_M$ is to solve the eigenvalue equations:
\beq\label{fuzlap}
\sfrac 1 {\l^2}[X_i,[X_i, \mathbb{Y^J_M}]]=J(J+1)~,\quad \sfrac 1{\l}[X_3,\mathbb{Y}^J_M]=M\mathbb{Y}^J_M~,\quad J=0,\dots, 2j,\;M=-J,\dots,J~.
\eeq       
Recalling that the vector space  $\text{Mat}_{2j+1} ({\C})$ as a tensor product naturally carries the action of the generators (\ref{gensu2a})-(\ref{gensu2c}) given as:
\beq \label{fuzder}
X_i \triangleright \Psi := [X_i,\Psi]~,\quad \forall \Psi \in \text{Mat}_{2j+1} ({\C})~, 
\eeq    
it is easy to see that solving Eqs. (\ref{fuzlap}) is equivalent to the decomposition
of $\text{Mat}_{2j+1} ({\C})$ into irreducible components given by: 
\beq 
D^j\otimes D^j= \bigoplus_{J=0}^{2j} D^J~.
\eeq
Consequently, the dimension of the space $\text{Mat}_{2j+1} ({\C})$ matches the dimension of the 
truncated space $L^2_{2j+1}(S^2)$ of square integrable functions on the sphere $S^2$ spanned by 
spherical harmonics $Y^J_M(\theta,\phi),\;J=0,\dots,2j,\;M=-j,\dots,j$. This enables to establish the following map:
\beq 
{\cal Q}: L^2_{2j+1}(S^2) \rightarrow \text{Mat}_{2j+1} ({\C}),\quad Y^J_M(\theta,\phi) \mapsto \mathbb{Y}^J_M~,
\eeq
which extends to an isomorphism of vector spaces due to the uniqueness of the corresponding expansions:
\beq 
{\cal Q}: f(\theta,\phi)=\sum_{J,M} f^J_M Y^J_M (\theta,\phi) \mapsto F=\sum_{J,M} F^J_M \mathbb{Y}^J_M~.  
\eeq
Having at hand such a map, one can introduce a noncommutative product of truncated functions on the sphere $S^2$ as:
\beq\label{prods2}
f_1\star f_2= {\cal Q}^{-1}\left({\cal Q}(f_1){\cal Q} (f_2)\right),
\eeq
which for practical purposes reduces merely to matrix multiplication.
Finally, one can show that in the limit $\lambda \to 0$ and $j\to \infty$, taken in a such way that 
$r=\lambda j$ is fixed, the isomorphism ${\cal Q}$ of vector spaces becomes an isomorphism of commutative algebras.
Particularly, the eigenvalue equations (\ref{fuzlap}) translate to eigenvalue equations for the Laplacian on the sphere and the third component of angular momentum. 
Accordingly, the constructed noncommutative algebra of functions   
together with the differential algebra determined by fuzzy derivatives (\ref{fuzder}) \cite{Madore:2000aq,Grosse:1993uq}, can be regarded as a quantized version of the sphere $S^2$, known as a fuzzy sphere
and denoted as $S^2_{\lambda,j}$.

Allowing $X_i$ to live in a reducible representation and keeping $\lambda$ fixed, a particular variant of noncommutative space, known as  $\mathbb{R}^{3}_{\lambda}$,
can be obtained as a direct sum of fuzzy spheres with all possible radii determined by $2j\in\N$ \cite{Hammou:2001cc,Wallet:2016ilh,Vitale:2012dz,Vitale:2014hca} 
\beq 
\mathbb{R}^{3}_{\lambda} = \sum_{2j\in\N} S^{2}_{\lambda, j}= \bigoplus_{2j\in\mathbb{N}} \textrm{Mat}_{2j+1} (\mathbb{C})~.
\eeq 
Thus $\mathbb{R}^{3}_{\lambda}$ can be thought of as a discrete foliation of 3D Euclidean space by multiple fuzzy 2-spheres, each being a leaf of the foliation \cite{DeBellis:2010sy}.
The structure and relation of $\mathbb{R}^{3}_{\lambda}$ to hermitian generators of $\mf{su}(2)$ in a matrix basis appears e.g. in \cite{Wallet:2016ilh}.
In that paper, the standard basis (\ref{gensu2a})-(\ref{gensu2c}) is considered for every $j$, so that $\R^3_{\l}$ inherits an orthogonal basis with respect to the inner product  $(a,b) = \text{Tr}(a^{\dagger}b),$
where the trace for any $a,b \in \R^{3}_{\lambda}$ is
\begin{equation}
\text{Tr}(ab) = 8\pi \lambda^{3} \sum_{2j \in \N} (2j+1) \text{tr}_{j}(A^{j}B^{j})~,
\end{equation}
with $A^{j}, B^{j} \in \text{Mat}_{2j+1} $. A dequantization map for $\mathbb{R}^{3}_{\lambda}$ 
can be constructed using coherent states and utilizing the fiber bundle structure of the group SU(2). 
The geometry underlying this construction can be understood following the discussion elaborated in \cite{Grosse:1993uq}. 
It is well known that as a smooth manifold, the group SU(2) is homeomorphic to the three-dimensional sphere $S^3$, which is a principal U(1) bundle over $S^2$.
The conventional orthogonal basis for the space of functions on the sphere $S^3$ is given by Wigner representation functions $D^{j}_{mn}(g)$ for SU(2), defined as matrix elements in an irreducible representation $D^j$.
Writing an element of SU(2) using Euler angles $(\psi,\theta,\phi)$ 
\bea
g(\psi,\theta,\phi)=e^{i\phi J_3} e^{i\theta J_1} e^{i\psi J_3}~,\quad g\in \text{SU}(2)~,
\eea
the functions $D^j_{mn}$ are given as: 
\bea
D^j_{mn}(g)=\langle jm|g(\psi,\theta,\phi)|jn\rangle=e^{im\psi}e^{in\phi} i^{m-n}P^j_{mn}(\cos \theta)~,\, m,n=-j,\dots,j\;, \, 2j\in\N~,
\eea
where the functions $P_{mn}^j$ are related to the finite hypergometric series $_2F_1$ \cite{vilenkin}.
They are orthogonal with respect to the Haar measure $\mu(g)$ on the group SU(2):
\bea  
\int \mu(g) D^j_{mn}(g) D^{j^\prime *}_{kl}(g)=\frac{1}{2j+1} \delta_{j j^\prime}\delta_{mk} \delta_{nl}~,
\eea
and therefore any square integrable function can be written as:
\bea\label{expand}
f(g)=\sum_{j=0}^\infty \sum_{m,n=-j}^{j} (2j+1) F^j_{mn}{D^{j}_{mn}}(g)~,
\eea
where the coefficients $F^j_{mn}$ are given by 
\bea\label{fcoe}
F^j_{mn}=\int \mu(g) f(g) {D^{j *}_{mn}}(g)~.
\eea
Collecting the coefficients $F^j_{mn}$ for fixed $j$ and organizing them into a matrix, one can interpret (\ref{fcoe}) as a map
from the space of functions on $S^3$ to the operators acting on the Hilbert space carrier of $D^j$. Such a map can be generalized 
for any representation $T(g)$ of any locally compact group $G$:
\bea\label{ga}
F=\int \mu(g) \tilde{f}(g) T(g)~,
\eea
where $\tilde{f}(g)$ is more appropriately interpreted as a distribution with compact support.
The space of operators (\ref{ga}) has the structure of an associative algebra with respect to matrix multiplication, known as a group algebra.
Alternatively, the group algebra can be realized as an algebra of functions with product defined by 
convolution:
\bea\label{con}
(\tilde{f}_1\diamond  \tilde{f}_2)(g)=\int \mu(g^\prime) \tilde{f}_1(g{g^\prime}^{-1})\tilde{f}_2(g^\prime)~,
\eea
which follows from the consistency condition:
\bea\label{convolalg}
F_1F_2=\int \mu(g) (\tilde{f}_1\diamond  \tilde{f}_2)(g) T(g)~.
\eea
Having at hand a certain homogeneous space $G/H$, with $H$ being the stabilizer subgroup of $G$, one
considers the group manifold as a principal $H$ bundle over $G/H$. Then using coherent states associated 
with the representation of $G$ induced by the one-dimensional representation of $H$, one can define a map from the group algebra
to the algebra of functions on $G/H$ with product which is in general noncommutative.
In the case of SU(2), the irreducible representations can be viewed as being induced by the one-dimensional representation of 
the one-parameter subgroup generated by $J_3$ so that the coset space SU(2)/U(1) is the sphere $S^2$.
Furthermore, using coherent states defined by 
\bea\label{chst}
|\theta \phi\rangle=\sum_{m=-j}^j \sqrt{\frac{(2j)!}{(j+m)!(j-m)!}} \left(-\sin{\frac{\theta}{2}}\right)^{j+m}
\left(\cos{\frac{\theta}{2}}\right)^{j-m} e^{-i(j+m)\phi} |jm\rangle~,
\eea
one assigns a function on the sphere $S^2$ to any operator given as in Eq. (\ref{ga}):
\bea
f(\theta,\phi)=\langle \theta \phi|F|\theta \phi\rangle~.
\eea   
The associative noncommutative product of functions on the sphere $S^2$, as defined by (\ref{prods2}), 
can be written as:  
\bea
(f_1 \star f_2)(\theta,\phi) =\langle \theta \phi|F_1 F_2|\theta \phi\rangle~.
\eea
The coherent states defined in (\ref{chst}) are standard SU(2) coherent states, built on the lowest weight state:
\bea 
|\theta \phi\rangle=D^j(g(0,\theta,\phi))|j -\!\!j\rangle~.
\eea 
These coherent states can be generalized considering  
unitary irreducible representations of SU(2) as being built on two copies of the Hilbert space of
the harmonic oscillator.
Following the Schwinger construction, one introduces usual creation and annihilation operators
$[a_\alpha,a^\dagger_\beta]=\delta_{\alpha\beta},\;\alpha,\beta=1,2$
and defines:
\bea
J_i=a^\dagger_\alpha \sigma_i^{\alpha\beta} a_\beta~,
\eea
where $\sigma^i$ are the Pauli matrices.
Then, a unitary irreducibile representation is given by the action on the subspace spanned by the states:
\bea\label{bas}
|jm\rangle=\frac{(a_1^\dagger)^{j+m}}{\sqrt{(j+m)!}}\frac{(a_2^\dagger)^{j-m}}{\sqrt{(j-m)!}}|0\rangle~,
\eea
with fixed $j$.
Inserting (\ref{bas}) into (\ref{chst}) and performing suitable transformations which relate equivalent representations \cite{vilenkin}, one can see 
that a simple generalization of (\ref{chst}) can be defined as:
\bea\label{genc}
|z_1z_2\rangle=\sum_{2j\in\N} \frac{e^{-\bar{z}z}}{\sqrt{2j!}}\sum_{m=-j}^j \sqrt{\frac{(2j)!}{(j+m)!(j-m)!}} z_1^{j+m}z_2^{j-m} |jm\rangle ~,
\eea
where any particular summand for fixed $j$ corresponds to (\ref{chst}). 
From this construction it is clear that the states (\ref{genc}) are related to the representation of SU(2) which is the 
direct sum of all possible irreducible representations. Consequently, the (de)quantization map and the noncommutative associative product 
for  $\mathbb{R}^{3}_{\lambda}$ can be defined as:
\bea
f(z_1,z_2)= \langle z_1 z_2|F| z_1 z_2\rangle,\;z_1,z_2\in \mathbb{C}~,\;
(f_1* f_2) (z_1,z_2)=\langle z_1z_2|F_1F_2| z_1z_2\rangle~.
\eea
Finally, we note that the representation space of $\mathbb{R}^{3}_{\lambda}$ can be safely truncated at some $j_ {max}$, which enables to 
interpret $\mathbb{R}^{3}_{\lambda}$ as a subalgebra of the matrix algebra $\text{Mat}_N ({\C})$, where  $N=(2j_{max}+1)(j_{max}+1)$. 
A basis of the space  $\text{Mat}_N ({\C})$ can be established in analogy to the aforementioned standard basis introducing elementary matrices 
$v^{jj^\prime}_{mn}, j,j^\prime=0,1/2,\cdots,j_{max},\;-j<m< j,\;-j^\prime< n< j^\prime$ for endomorphisms $|jm\rangle\langle j^\prime n|$. 
Defining the projector $P_j=\sum_{m=-j}^j v^{jj}_{mm}$ on the particular fuzzy sphere contained in $\text{Mat}_N ({\C})$,
as explained in \cite{Wallet:2016ilh}, one can introduce the radius operator $X_{0}$ which generates the center of $\R^{3}_{\lambda}$: 
\begin{equation}
X_{0} =  \lambda \sum_{ 2j \in \N} j(j+1)P_{j}~.
\end{equation}
It is related to the quadratic Casimir of SU(2) by:
\begin{equation}
X_{0}^{2} +\l X_0= \sum_{l=1}^{3} X_{l}^{2}~.
\end{equation}
From this point of view, $\mathbb{R}^{3}_{\lambda}$ is the subalgebra of  $\text{Mat}_N ({\C})$ 
with elements subject to the constraint $[X_0,\Psi]=0 \Leftrightarrow \Psi \in \mathbb{R}^{3}_{\lambda}$, which by the dequantization map induces the constraint on functions \cite{Hammou:2001cc}
\bea
(\bar{z}_\alpha\bar{\partial}_\alpha-z_\alpha\partial_\alpha)f(z_1,z_2)=0~,
\eea 
considered as functions on $\mathbb{C}^2$.
Our last and for our purpose the most important note related to $\mathbb{R}^{3}_{\lambda}$ is that due to the relation to the group algebra of SU(2) this space 
caries a natural action of SO(4) induced by left and right translations in the group.

As presented, the construction in the Euclidean case has a direct analog in the case of Minkowski signature. In the same way as fuzzy 2-spheres act as a backdrop in the construction of the 3D fuzzy space based on 
SU(2), a set of fuzzy dS$_2$ of different radii may be used to define a 3D fuzzy space based on SU(1,1), which is the two-fold cover of SO(2,1) in the same way as SU(2) covers SO(3). The commutation relations of the algebra are{\footnote{We use the same notation for the generators of $\mf{su}(2)$ and $\mf{su}(1,1)$.}}
\be \label{su11}
[J_1,J_2]=-iJ_3~,\quad [J_2,J_3]=iJ_1~,\quad [J_3,J_1]=iJ_2~.
\ee

According to Ref. \cite{Jurman:2013ota}, one can consider three operators $X_i=\l J_i$ satisfying
\be
[X_i,X_j]=i\l C_{ij}{}^{k}X_k~,
\ee
where now $C_{ij}{}^{k}$ are the structure constants of $\mf{su}(1,1)$, read off from the commutators \eqref{su11}. Imposing the Casimir relation
\be \label{casimirads}
\sum_{i,j}\eta_{ij}X_iX_j=\l^2 j(j-1)~,
\ee
where $\eta_{ij}$ is the 3D Minkowski metric, one obtains a fuzzy hyperboloid, provided that the irreducible representations of $\mf{su}(1,1)$ are appropriately chosen. The main difference to the previous case is that now the corresponding group is non-compact and therefore it has no finite-dimensional unitary irreducible representations. However, it naturally possesses infinite-dimensional ones. It was argued in \cite{Jurman:2013ota} that the relevant irreducible representations for the construction of a fuzzy hyperboloid  are chosen from the principal continuous series.

For the purposes of the present paper, and in full analogy to the Euclidean case,  we relax the condition
\eqref{casimirads}. Thus the three operators $X_i$ live in infinite-dimensional reducible representations instead, taking a block-diagonal form with each block being a fuzzy hyperboloid. This is a  foliation with leaves being fuzzy hyperboloids of different radii. More details on the full construction of this 3D fuzzy space, its algebra of functions and its differential calculus will be given elsewhere.

\section{Gravity as gauge theory on 3D fuzzy spaces}
\subsection{The Lorentzian case}
Our proposal here is that in the same spirit as the gauging of the Poincar\'e/(A)dS algebra in the classical case, the covariant coordinate on a fuzzy 3D space, such as the ones discussed in Section \ref{sec4}, should accommodate the information of a noncommutative vielbein and spin connection.{\footnote{A similar idea was pursued in Refs. \cite{Nair:2001kr,Abe:2002in,Nair:2006qg}, however our construction here is different.}}

We wish to consider the 3D case with positive cosmological constant.
Thus the relevant isometry groups are $\text{SO}(3,1)$ in the Lorentzian case and $\text{SO}(4)$ in the Euclidean case. Since our plan is to write down a non-Abelian noncommutative gauge theory, it is imperative that the issue mentioned at the end of Section \ref{sec3} is treated carefully. Our approach will be inspired by the analogous one of Ref. \cite{Aschieri1} in the Moyal-Weyl case. The difference is that the group structures considered there refer to four dimensions without cosmological constant, while here we consider three dimensions with cosmological constant.

Thus we first consider the corresponding spin groups. They are isomorphic to $\text{Spin}(3,1)=\text{SL}(2;\C)$ and
$\text{Spin}(4)=\text{SU}(2)\times \text{SU}(2)$. However, the anticommutators of generators in these groups do not close. This is taken care of as follows. For the first case, we focus on the spinor
representation, which is generated by the elements $\S_{AB}=\sfrac 12\g_{AB}=\sfrac 14 [\g_A,\g_B], A=1,2,3,4$, $\gamma_A$ being 4D Lorentzian gamma matrices.
Due to the product relation \cite{VanProeyen:1999ni}
\be
\g_{AB}\g^{CD}=2\d^{[C}_{[B}\d^{D]}_{A]}+4\d^{[C}_{[B}\g_{A]}{}^{D]}+i\varepsilon_{AB}{}^{CD}\g_5~,
\ee
one finds the commutation and anticommutation relations
\bea
[\gamma_{AB},\gamma_{CD}] &=& 8\eta_{[A[C}\gamma_{D]B]}~,\\
\{\gamma_{AB},\gamma_{CD}\} &=& 4\eta_{C[B}\eta_{A]D} {\one} +2i\epsilon_{ABCD}\gamma_5\,.\label{anti}
\eea
Thus, due to the second relation \eqref{anti}, it turns out that $\g_5=i\g_1\g_2\g_3\g_4$ and the identity have to be included in the algebra.
Therefore we extend the algebra by these two elements, which leads to an 8-dimensional algebra; in fact this is nothing but
the extension of $\text{SL}(2;\C)$ to $\text{GL}(2;\C)$, generated by\footnote{We use the same set of $\gamma_4$-hermitian generators as in Ref. \cite{Aschieri1} (in our conventions, $\gamma_4$ corresponds to the $\g_0$ of that paper, and our metric signature is the opposite one, with $\eta_{44}=-1$); see also \cite{Aschieri:2012vf}, Sec. 4.2, for a detailed explanation.} $\{\g_{AB},\g_5,i\one\}$. In the second case, one has to extend the $\text{SU}(2)\times \text{SU}(2)$ symmetry to $\text{U}(2)\times \text{U}(2)$. Here we will discuss in detail the first case, and comment on the second case at the end.

In $\text{SO}(3)$ notation,
we have the generators $\g_{ab}$ and $\g_a=\g_{a4}$ with $a=1,2,3$. For notational simplicity we can also define
$\widetilde\g^a=\epsilon^{abc}\g_{bc}$. It is useful to write the commutation and anticommutation relations for $\gamma$'s and $\widetilde{\g}$'s, since they will both be used in what follows. They are
\bea
&&[\tilde{\gamma}^a,\tilde{\gamma}^b]=-4\epsilon^{abc}\tilde{\gamma}_c\,, \,
[\gamma_a,\tilde{\gamma}_b]=-4\epsilon_{abc}\gamma^c\,, \, [\gamma_a,\gamma_b]=\epsilon_{abc}\tilde{\gamma}^c\,,\\
&&\{\tilde{\gamma}^a,\tilde{\gamma}^b\}=-8\eta^{ab}{\one}\,,\, \{\gamma_a,\tilde{\gamma}^b\}=4i\delta_a^b\gamma_5\,,\, \{\gamma_a,\gamma_b\}=2\eta_{ab}{\one}\,, \\
&&  [\gamma^5,\gamma^{AB}]=0\,,\, \{\gamma^5,\gamma^{AB}\}=i\epsilon^{ABCD}\g_{CD}~,\, \{\g_a,\g_5\}=i\tilde{\g}_a~,\, \{\tilde{\g}_a,\g_5\}=-4i\g_a~.
\eea

According to the above, we consider $\text{GL}(2;\C)$ as the gauge group and we identify noncommutative coordinates $X_{a}$ with the three operators of the 3D fuzzy space discussed in Section \ref{sec4}. Following the discussion of Section \ref{sec3}, the covariant coordinates we consider are
\be
{\cal X}_{\mu}=\d_{\m}{}^aX_a+{\cal A}_{\m}~,
\ee
where ${\cal A}_{\m}={\cal A}_{\mu}^{\bar{a}}(X_a)\otimes T^{\bar{a}}$ are $\text{GL}(2;\C)$-valued gauge fields.
Note that the component fields are not any longer functions on a classical manifold, but instead they are operator-valued, which explains the tensor product structure. According to the discussion on the $\text{GL}(2;\C)$ generators, we expand the gauge field as follows,
\be
{\cal A}_{\m}=e_{\m}{}^a(X)\otimes \gamma_a+ \omega_{\m}{}^{a}(X)\otimes \widetilde\g_{a}+{A}_{\mu}(X)\otimes i\one+\widetilde{A}_{\m}(X)\otimes \g_5~.
\ee
A similar expansion holds for the gauge parameter:
\be
\epsilon=\xi^a(X)\otimes \gamma_a+ \l^{a}(X)\otimes \widetilde\g_{a}+\epsilon_0(X)\otimes i\one+\widetilde\epsilon_0(X)\otimes\g_5~.
\ee

Using the general form of the covariant transformation rule{\footnote{More precisely, due to the choice of GL(2,$\C$) generators, which includes $i\one$, we use here $\d{\cal X}=[\epsilon,{\cal X}]$.}} \eqref{gentrafo}, we can find the transformations of the component fields, as in the commutative case. The main difference here is that one has to pay attention to the order of fields, in other words to use the formula \eqref{nonAbelianCR}. The
transformations turn out to be (denoting $X_{\mu}=\d_{\m}{}^aX_a$)
\bea
\delta e_\mu^{~a}&=& -i[X_{\m}+A_\mu,\xi^a]-2\{\xi_b,\omega_{\mu c}\}\epsilon^{abc}
						-2\{\lambda_b,e_{\mu c}\}\epsilon^{abc}+i[\epsilon_0,e_\mu^{~a}]-\nn\\ 
						&&-2i[\l^a,\widetilde{A}_{\mu}]-2i[\widetilde{\epsilon}_0,\omega_{\mu}{}^a]~,\\
\delta\omega_\mu^{~a}&=& -i[X_{\m}+ A_\mu,\lambda^a]
							+\sfrac{1}{2}\{\xi_b,e_{\mu c}\}\epsilon^{abc}-2\{\lambda_b,\omega_{\mu c}\}\epsilon^{abc}+i[\epsilon_0,\omega_\mu^{~a}]+\nn\\ 
							&&+\sfrac i2 [\xi^a,\widetilde{A}_{\mu}]+\sfrac i2[\widetilde{\epsilon}_0,e_{\mu}{}^a]~,\\
\delta {A}_\mu&=&-i[X_{\m}+{A}_\mu,\epsilon_0]
						-i[\xi_a,e_\mu^{~a}]+4i[\lambda_a,\omega_\mu^{~a}]
						-i[\tilde{\epsilon}_0,\widetilde{A}_\mu]~,\label{deltaA}\\
\delta\widetilde{A}_\mu&=&-i[X_{\m}+{A}_\mu,\tilde{\epsilon}_0]
						+ 2i[\xi_a,\omega_\mu^{~a}]+2i[\lambda_a,e_\mu^{~a}]+i[\epsilon_0,\widetilde{A}_\mu]~.
\eea
Let us pause here to comment on these transformation rules. First, had we not considered a non-Abelian gauge group, we would have obtained just an Abelian gauge theory on the 3D fuzzy space. Indeed, this effectively amounts to setting $e_{\mu}{}^a=\o_{\mu}{}^a=0$ and $\widetilde{A}_{\mu}=0$, the only gauge parameter being $\epsilon_0$, in which case only Eq. \eqref{deltaA} is not trivial and it becomes
$$
\d A_{\mu}=-i[X_{\mu},\epsilon_0]+i[\e_0,A_{\mu}]~,
$$
which is the expected transformation rule for a noncommutative Maxwell gauge field. Thus we observe that
the Maxwell sector is always there, whether or not the dreibein is trivial, $X_{\mu}+{A}_\mu$ being the corresponding covariant coordinate.
Second, in the naive commutative limit, where the Yang-Mills and gravity fields disentangle and we can set $A=0$, the inner derivation becomes $[{X}_\mu, f]\to -i\partial_\mu f$. Thus in this limit we obtain the following transformations
for the dreibein and spin connection,
\bea\label{cl}
\delta e_\mu^{~a}&=&-\partial_\mu\xi^a-4\xi_b\omega_{\mu c}\epsilon^{abc}-4\lambda_be_{\mu c}\epsilon^{abc}~,\\
\delta\omega_\mu^{~a}&=&-\partial_\mu\lambda^a +\xi_be_{\mu c}\epsilon^{abc}-4\lambda_b\omega_{\mu c}\epsilon^{abc}~.
\eea
It is then  observed that using the redefinitions
$\gamma_a\to \sfrac{2i}{\sqrt{\Lambda}} P_a~, \tilde\gamma_a\to -4J_a~,$ and also
 $4\lambda^a\to \lambda^a,\;\xi^a\sfrac{2i}{\sqrt\Lambda}\to -\xi^a,\;e^a_\mu\rightarrow\sfrac{\sqrt\Lambda}{2i}e^a_\mu, \;\omega^a_\mu\rightarrow -\sfrac{1}{4}\omega^a_\mu$, these transformation rules are identical to Eqs. \eqref{deltae} and \eqref{deltao}. Thus in the commutative limit, the transformations of \cite{WittenCS} for the fields of three-dimensional gravity are recovered. Of course, in order to really state this, we also have to show that the 3D fuzzy space has a meaningful commutative limit itself.

Next we calculate the commutator of the covariant coordinates in order to obtain the curvature tensors.
As explained before, since we are dealing with a case where the right-hand side of the commutators in the algebra are linear in generators, an additional linear term is included in the definition of curvature,
which reads
\bea
\mathcal{R}_{\mu\nu}&=&[\mathcal{X}_\mu,\mathcal{X}_\nu]-i\l C_{\mu\nu}{}^{\rho}\mathcal{X}_\rho
~. \eea
The curvature tensor can be expanded in the generators of GL(2,$\C$) as:
\begin{equation}
\mathcal{R}_{\mu\nu}=T^a_{\mu\nu}(X)\otimes \gamma_a+R_{\mu\nu}^a(X) \otimes \tilde\gamma_a+F_{\mu\nu}(X)\otimes i\one+\widetilde{F}_{\mu\nu}(X)\otimes\gamma_5\,.
\end{equation}
Therefore, we obtain the following expressions for the various tensors:
\begin{align}
T^a_{\mu\nu}&=i[X_{\m}+{A}_\mu,e_\nu^{~a}]-i[X_{\nu}+{A}_\nu,e_\mu^{~a}]-2\{e_{\mu b},
				\omega_{\nu c}\}\epsilon^{abc}-2\{\omega_{\mu b},e_{\nu c}\}\epsilon^{abc}-\nn\\ 
				&\quad -2i[\omega_{\mu}{}^a,\widetilde{A}_{\nu}]+2i[\omega_{\nu}{}^a,\widetilde{A}_{\mu}]-i\l C_{\mu\nu}^{~~~\rho}e^{~a}_\rho~,
\\
R^a_{\mu\nu}&=i[X_{\m}+{A}_\mu,\omega_\nu^{~a}]-i[X_{\nu}+{A}_\nu,\omega_\mu^{~a}]
			-2\{\omega_{\mu b},\omega_{\nu c}\}\epsilon^{abc}+\sfrac{1}{2}\{e_{\mu b},e_{\nu c}\}\epsilon^{abc}+\nn\\ & +\sfrac i2[e_{\mu}{}^a,\widetilde{A}_{\nu}]-\sfrac i2[e_{\nu}{}^a,\widetilde{A}_{\mu}]-i\l C_{\mu\nu}^{~~~\rho}\omega_\rho^{~a}~,\\
F_{\mu\nu}&=i[X_{\m}+{A}_\mu,X_{\nu}+{A}_\nu]-i[e_\mu^{~a},e_{\nu a}]+
			4i[\omega_\mu^{~a},\omega_{\nu  a}]-i[\widetilde{A}_\mu,\widetilde{A}_\nu]-i\l C_{\mu\nu}^{~~~\rho}(X_{\rho}+{A}_\rho)~,\\
\widetilde{F}_{\mu\nu}&=i[X_{\m}+{A}_\mu,\widetilde{A}_\nu]-i[X_{\nu}+{A}_\nu,\widetilde{A}_\mu]+2i[e_\mu^{~a},\omega_{\nu a}]+2i[\omega_\mu^{~a},e_{\nu a}]-i\l C_{\mu\nu}^{~~~\rho}\widetilde{A}_\rho~.
\end{align}
Once more, the  commutative limit coincides with the expected result appearing in Eqs.
\eqref{tmn} and \eqref{rmn}, using the aforementioned redefinitions.

\subsection{The Euclidean case} 
As explained in the beginning of the present section,
in the Euclidean case one has to work with the gauge group $\text{U}(2)\times \text{U}(2)$ in fixed representation. Recalling that each U(2) is spanned by four generators given by the Pauli matrices and the unit matrix, this
means that the expansions of the gauge field and the gauge parameter should involve the 4$\times$4 matrices
\bea
J^L_a=\begin{pmatrix}
	\sigma_a & 0 \\ 0 & 0
\end{pmatrix}~, \quad J^R_a=\begin{pmatrix}
0 & 0 \\ 0 & \sigma_a
\end{pmatrix}~,
\eea
and
\bea
J^L_0=\begin{pmatrix}
	\one & 0 \\ 0 & 0
\end{pmatrix}~, \quad J^R_0=\begin{pmatrix}
	0 & 0 \\ 0 & \one
\end{pmatrix}~.
\eea
However, one should be careful in identifying what the noncommutative vielbein and spin connection are in the expansion of the gauge field. In order to achieve the correct interpretation, we consider
\bea
P_a=\sfrac 12 (J_a^L-J_a^R)=\frac 12 \begin{pmatrix}
	\sigma_a & 0 \\ 0 & -\sigma_a
\end{pmatrix}~, \quad M_a=\sfrac 12 (J_a^L+J_a^R)=\frac 12 \begin{pmatrix}
\sigma_a & 0 \\ 0 & \sigma_a
\end{pmatrix}~,
\eea
and also
\bea
\one=J_0^L+J_0^R~, \quad \gamma_5=J_0^L-J_0^R~.
\eea
These indeed satisfy the expected commutation and anticommutation relations,
\bea
&& [P_a,P_b]=i\epsilon_{abc}M_c~, \quad [P_a,M_b]=i\epsilon_{abc}P_c~, \quad [M_a,M_b]=i\epsilon_{abc}M_c~,
\\
&& \{P_a,P_b\}=\sfrac 12 \delta_{ab}\one~,\quad \{P_a,M_b\}=\sfrac 12 \delta_{ab}\g_5~, \quad \{M_a,M_b\}=\sfrac 12 \d_{ab}\one~.
\\
&& [\gamma_5 , P_a] = [\gamma_5, M_a] = 0~,\quad
\{\gamma_5 , P_a\} =2 M_a~,\quad  \{\gamma_5, M_a\} = 2 P_a~.
\eea
One then proceeds as before, with the covariant coordinate
\be
{\cal X}_{\mu}= X_{\mu}\otimes i\one +e_{\mu}{}â\otimes P_a+\omega_{\mu}{}â\otimes M_a+A_{\mu}\otimes i\one+{\widetilde{A}}_{\mu}\otimes \g_5~~,
\ee
and the gauge parameter
\be
\epsilon=\xi^a\otimes P_a+\l^a\otimes M_a+\epsilon_0\otimes i\one+\widetilde{\epsilon}_0\otimes\g_5~.
\ee
The only difference to the previous case is the metric signature, thus we do not repeat the formulas here.

\section{Action of 3D fuzzy gravity}
As a final step, we would like to write down an action incorporating the above curvatures. First, in Ref. \cite{Jurman:2013ota} it was shown that fuzzy 2-hyperboloids provide dynamical brane solutions of a Yang-Mills type matrix model with its characteristic square commutator term. From our viewpoint, working in 3D and recalling that in this number of dimensions general relativity has no dynamics, we propose the following action (cf. \cite{Gere:2013uaa}){\footnote{A similar action was proposed in Ref. \cite{Valtancoli:2003ve} for a gravity theory on the fuzzy sphere. See also \cite{Alekseev:2000fd}.}}
\bea S_0= \sfrac 1 {g^2} \text{Tr}\left(\sfrac i3C^{\mu\nu\rho}{ X}_\mu{ X}_\nu{X}_\rho
-m^2{ X}_{\mu}{ X}^{\mu} \right)~. \eea
The 3D fuzzy space we considered is indeed a solution of the field equations derived from this action,
\be 
[X_{\mu},X_{\nu}]-2im^2C_{\mu\nu}{}^{\rho}X_{\rho}=0~,
\ee
 when $2m^2=\l$.

Furthermore, we would like to write the action including the gauge fields. One could either consider the fluctuations around the above solution, or directly write down an action for the curvatures in the spirit of \cite{Madore:2000en}.
This action should be written in terms of the covariant coordinates ${\cal X}_{\mu}$ and it should also
contain a prescription for taking the trace over the gauge algebra. Regarding this matter, although there are
two different trace prescriptions available \cite{WittenCS}, only one of them works in our case. This is
because we have fixed the representation and used gamma matrices in our expansions. Thus the
prescription imposed on us by the algebra of gamma matrices is the one corresponding to Eq. \eqref{tr2}.
More specifically, we use the trace relations
\bea
\text{tr}{\left(\gamma_a\gamma_b\right)}= 4\eta_{ab}~,\quad \text{tr}{\left(\tilde\gamma_a\tilde\gamma_b\right)}= -16\eta_{ab}~. \eea
Then the action we propose is
\bea S= \sfrac 1{g^2}\text{Tr}~\text{tr}\left(\sfrac i3C^{\mu\nu\rho}{\cal X}_\mu{\cal X}_\nu{\cal X}_\rho
-\sfrac {\l}{2}{\cal X}_{\mu}{\cal X}^{\mu} \right)
~, \eea
where the first trace $\text{Tr}$ is over the matrices $X$ and the second trace $\text{tr}$ is over the algebra.
We can rewrite this action as
\be
S=\sfrac 1{6g^2} \text{Tr}~\text{tr}\left(iC^{\mu\nu\rho}{\cal X}_\mu {\cal R}_{\nu\rho}\right)+S_{\l}~,
\ee
where $S_{\l}=-\sfrac {\l}{6g^2}\text{Tr}~\text{tr}\left({\cal X}^{\m}{\cal X}_{\m}\right)$ and it vanishes in the limit $\l \to 0$.
Using the explicit form of the algebra trace, the first term in the action is proportional to
\bea
\text{Tr}~C^{\mu\nu\rho}(e_{\mu a}T^a_{\nu\rho}-4\omega_{\mu a}R^a_{\nu\rho}-(X_{\mu}+{ A}_\mu) F_{\nu\rho}+\widetilde A_\mu\widetilde F_{\nu\rho})~.\eea
This action is similar to the one obtained in Section 2.3 of Ref. \cite{WittenCS}. Upon taking the
commutative limit and performing a field redefinition, the first two terms are identical to that action; however, in the present case we necessarily obtain an additional sector, associated to the additional gauge fields that cannot
decouple in the noncommutative case.

\section{Conclusions}

Based on known relations between gravity and gauge theory, we examined 3D noncommutative gravity with cosmological constant from the point of view of noncommutative gauge theory.
Our approach follows the standard path for constructing gauge theories on noncommutative spaces in terms of covariant coordinates. In doing so, we had to account for two issues: (a) What is the 3D noncommutative space on which we construct the gauge theory, and (b) what is the gauge group of the theory. Regarding the first, we considered 3D spaces with certain symmetries acting on them.
More specifically, for the Euclidean case we considered a discrete foliation of 3D space by fuzzy 2-spheres, based on reducible representations of SU(2), with the action of SO(4) on it. Similarly, in the Lorentzian case we considered a foliation by fuzzy 2-hyperboloids, yielding a 3D fuzzy space with
an SO(3,1) action.

The next step was to consider gauge theories on the above fuzzy spaces. Given that the symmetries acting on them are non-Abelian, one has to take care of the typical issue of non-Abelian noncommutative gauge theories and find in which algebra the gauge field takes values. This is solved by taking the corresponding double cover of SO(4) and SO(3,1) respectively and fixing the representation such that all anticommutators close. In the first case, one is led to U(2)$\times$U(2), while in the second case to GL(2;$\C$). We note that GL(2;$\C$) gauge theories for gravity have been considered before in \cite{Chamseddine:2003we,Aschieri1}, albeit in the context of 4D noncommutative gravity on Moyal-Weyl space without cosmological constant. This coincidence is perfectly reasonable, since 3D gravity with cosmological constant and 4D without one share the same underlying symmetry.

Having addressed the above two issues, we considered the corresponding gauge theories by writing the covariant coordinate and expanding the gauge fields and parameters in the generators of the algebra, identifying a noncommutative dreibein, a spin connection and two additional Maxwell fields. This led to the derivation of the transformation rules for these fields, which reproduce the standard ones in the
commutative limit. In addition, the corresponding curvatures were determined and a matrix action was proposed, which is related to one of the actions proposed in \cite{WittenCS} in the commutative limit.

 \paragraph{Acknowledgements.}
 We would like to thank P.~Aschieri, L.~Castellani, ~A.~H.~Chamseddine and D.~L\"ust for useful discussions. 
The work of A.Ch., L.J. and D.J. was supported by the Croatian Science Foundation under the project
IP-2014-09-3258 and by the H2020 Twinning Project No. 692194 "RBI-T-WINNING". We acknowledge support by the COST action QSPACE MP1405. G.Z thanks the MPI Munich for hospitality and the A.v.Humboldt Foundation for support. 
A.Ch. thanks LMU and MPI Munich for hospitality during a visit.


\begin{thebibliography}{99}
 \addtolength{\itemsep}{-4pt}



\bibitem{Utiyama:1956sy}
  R.~Utiyama,
  Phys.\ Rev.\  {\bf 101} (1956) 1597.
   doi:10.1103/PhysRev.101.1597

\bibitem{Kibble:1961ba}
  T.~W.~B.~Kibble,
  J.\ Math.\ Phys.\  {\bf 2} (1961) 212.
  doi:10.1063/1.1703702

\bibitem{MacDowell:1977jt}
  S.~W.~MacDowell and F.~Mansouri,
  Phys.\ Rev.\ Lett.\  {\bf 38} (1977) 739
   Erratum: [Phys.\ Rev.\ Lett.\  {\bf 38} (1977) 1376].
 doi:10.1103/PhysRevLett.38.1376, 10.1103/PhysRevLett.38.739

\bibitem{Kibble:1985sn}
  T.~W.~B.~Kibble and K.~S.~Stelle,
  In Ezawa, H. ( Ed.), Kamefuchi, S. ( Ed.): Progress In Quantum Field Theory, 57-81.

\bibitem{AchucarroTownsend}
A.~Achucarro and P.~K.~Townsend,
Phys.\ Lett.\ B {\bf 180} (1986) 89.
doi:10.1016/0370-2693(86)90140-1

\bibitem{WittenCS}
E.~Witten,
Nucl.\ Phys.\ B {\bf 311} (1988) 46.
doi:10.1016/0550-3213(88)90143-5

\bibitem{Madore:2000en}
J.~Madore, S.~Schraml, P.~Schupp and J.~Wess,
Eur.\ Phys.\ J.\ C {\bf 16} (2000) 161 
doi:10.1007/s100520050012
 [hep-th/0001203].

\bibitem{Chamseddine:2000si}
A.~H.~Chamseddine,
Phys.\ Lett.\ B {\bf 504} (2001) 33 
doi:10.1016/S0370-2693(01)00272-6
[hep-th/0009153].

\bibitem{Chamseddine:2003we}
A.~H.~Chamseddine,
Phys.\ Rev.\ D {\bf 69} (2004) 024015
doi:10.1103/PhysRevD.69.024015
[hep-th/0309166].

\bibitem{Aschieri1}
P.~Aschieri and L.~Castellani,
JHEP {\bf 0906} (2009) 086
doi:10.1088/1126-6708/2009/06/086
[arXiv:0902.3817 [hep-th]].

\bibitem{Aschieri2}
P.~Aschieri and L.~Castellani,
JHEP {\bf 0906} (2009) 087
doi:10.1088/1126-6708/2009/06/087
[arXiv:0902.3823 [hep-th]].

\bibitem{Ciric:2016isg}
M.~Dimitrijevi\'c \'Ciri\'c, B.~Nikoli\'c and V.~Radovanovi\'c,
Phys.\ Rev.\ D {\bf 96} (2017) no.6,  064029
doi:10.1103/PhysRevD.96.064029
[arXiv:1612.00768 [hep-th]].


\bibitem{Cacciatori:2002gq}
S.~Cacciatori, D.~Klemm, L.~Martucci and D.~Zanon,
Phys.\ Lett.\ B {\bf 536} (2002) 101
doi:10.1016/S0370-2693(02)01823-3
[hep-th/0201103].

\bibitem{Cacciatori:2002ib}
S.~Cacciatori, A.~H.~Chamseddine, D.~Klemm, L.~Martucci, W.~A.~Sabra and D.~Zanon,
Class.\ Quant.\ Grav.\  {\bf 19} (2002) 4029 
doi:10.1088/0264-9381/19/15/310
[hep-th/0203038].

\bibitem{Aschieri3}
P.~Aschieri and L.~Castellani,
JHEP {\bf 1411} (2014) 103
doi:10.1007/JHEP11(2014)103 
[arXiv:1406.4896 [hep-th]].

\bibitem{Banados:2001xw}
  M.~Banados, O.~Chandia, N.~E.~Grandi, F.~A.~Schaposnik and G.~A.~Silva,
  Phys.\ Rev.\ D {\bf 64} (2001) 084012
  doi:10.1103/PhysRevD.64.084012
  [hep-th/0104264].

\bibitem{Seiberg:1999vs}
N.~Seiberg and E.~Witten,
JHEP {\bf 9909} (1999) 032
doi:10.1088/1126-6708/1999/09/032
[hep-th/9908142].

\bibitem{Banks:1996vh}
T.~Banks, W.~Fischler, S.~H.~Shenker and L.~Susskind,
Phys.\ Rev.\ D {\bf 55} (1997) 5112
doi:10.1103/PhysRevD.55.5112
[hep-th/9610043].

\bibitem{Ishibashi:1996xs}
N.~Ishibashi, H.~Kawai, Y.~Kitazawa and A.~Tsuchiya,
Nucl.\ Phys.\ B {\bf 498} (1997) 467
doi:10.1016/S0550-3213(97)00290-3
[hep-th/9612115].

\bibitem{Aoki:1998vn}
H.~Aoki, S.~Iso, H.~Kawai, Y.~Kitazawa and T.~Tada,
Prog.\ Theor.\ Phys.\  {\bf 99} (1998) 713
doi:10.1143/PTP.99.713
[hep-th/9802085].

\bibitem{Hanada:2005vr}
M.~Hanada, H.~Kawai and Y.~Kimura,
Prog.\ Theor.\ Phys.\  {\bf 114} (2006) 1295
doi:10.1143/PTP.114.1295
 [hep-th/0508211].

\bibitem{Furuta:2006kk}
K.~Furuta, M.~Hanada, H.~Kawai and Y.~Kimura,
Nucl.\ Phys.\ B {\bf 767} (2007) 82
doi:10.1016/j.nuclphysb.2007.01.003
 [hep-th/0611093].

\bibitem{Yang:2006dk}
H.~S.~Yang,
Int.\ J.\ Mod.\ Phys.\ A {\bf 24} (2009) 4473
doi:10.1142/S0217751X0904587X
[hep-th/0611174].

\bibitem{Steinacker:2010rh}
H.~Steinacker,
Class.\ Quant.\ Grav.\  {\bf 27} (2010) 133001
doi:10.1088/0264-9381/27/13/133001
[arXiv:1003.4134 [hep-th]].

\bibitem{Kim:2011cr}
S.~W.~Kim, J.~Nishimura and A.~Tsuchiya,
Phys.\ Rev.\ Lett.\  {\bf 108} (2012) 011601
doi:10.1103/PhysRevLett.108.011601
[arXiv:1108.1540 [hep-th]].

\bibitem{Nishimura:2012xs}
J.~Nishimura,
PTEP {\bf 2012} (2012) 01A101
doi:10.1093/ptep/pts004
[arXiv:1205.6870 [hep-lat]].

\bibitem{Nair:2001kr}
V.~P.~Nair,
Nucl.\ Phys.\ B {\bf 651} (2003) 313
doi:10.1016/S0550-3213(02)01061-1
[hep-th/0112114].

\bibitem{Abe:2002in}
Y.~Abe and V.~P.~Nair,
Phys.\ Rev.\ D {\bf 68} (2003) 025002
doi:10.1103/PhysRevD.68.025002
[hep-th/0212270].

\bibitem{Valtancoli:2003ve}
P.~Valtancoli,
Int.\ J.\ Mod.\ Phys.\ A {\bf 19} (2004) 361
doi:10.1142/S0217751X04017598
[hep-th/0306065].

\bibitem{Nair:2006qg}
V.~P.~Nair,
Nucl.\ Phys.\ B {\bf 750} (2006) 321
doi:10.1016/j.nuclphysb.2006.06.009
[hep-th/0605008].

\bibitem{Buric:2006di}
M.~Buri\'c, T.~Grammatikopoulos, J.~Madore and G.~Zoupanos,
JHEP {\bf 0604} (2006) 054
doi:10.1088/1126-6708/2006/04/054
[hep-th/0603044].

\bibitem{Buric:2007zx}
M.~Buri\'c, J.~Madore and G.~Zoupanos,
SIGMA {\bf 3} (2007) 125
doi:10.3842/SIGMA.2007.125
[arXiv:0712.4024 [hep-th]].

\bibitem{Buric:2007hb}
M.~Buri\'c, J.~Madore and G.~Zoupanos,
Eur.\ Phys.\ J.\ C {\bf 55} (2008) 489
doi:10.1140/epjc/s10052-008-0602-x
[arXiv:0709.3159 [hep-th]].


\bibitem{Aschieri:2003vyAschieri:2004vhAschieri:2005wm}
  P.~Aschieri, J.~Madore, P.~Manousselis and G.~Zoupanos,
  JHEP {\bf 0404} (2004) 034
  doi:10.1088/1126-6708/2004/04/034
  [hep-th/0310072];
  ibid,  
  Fortsch.\ Phys.\  {\bf 52} (2004) 718
  doi:10.1002/prop.200410168
  [hep-th/0401200];
  ibid,
  hep-th/0503039.

\bibitem{Aschieri:2006uwSteinacker:2007ay}
  P.~Aschieri, T.~Grammatikopoulos, H.~Steinacker and G.~Zoupanos,
  JHEP {\bf 0609} (2006) 026
  doi:10.1088/1126-6708/2006/09/026
  [hep-th/0606021]; 
  H.~Steinacker and G.~Zoupanos,
  JHEP {\bf 0709} (2007) 017
  doi:10.1088/1126-6708/2007/09/017
  [arXiv:0706.0398 [hep-th]]; 


\bibitem{Lukierski:1991pn}
J.~Lukierski, H.~Ruegg, A.~Nowicki and V.~N.~Tolstoi,
Phys.\ Lett.\ B {\bf 264} (1991) 331.
doi:10.1016/0370-2693(91)90358-W

\bibitem{Lukierski:1992dt}
J.~Lukierski, A.~Nowicki and H.~Ruegg,
Phys.\ Lett.\ B {\bf 293} (1992) 344.
doi:10.1016/0370-2693(92)90894-A

\bibitem{Kim:2011ts}
S.~W.~Kim, J.~Nishimura and A.~Tsuchiya,
Phys.\ Rev.\ D {\bf 86} (2012) 027901
doi:10.1103/PhysRevD.86.027901
[arXiv:1110.4803 [hep-th]].

\bibitem{Snyder:1946qz}
H.~S.~Snyder,
Phys.\ Rev.\  {\bf 71} (1947) 38.
doi:10.1103/PhysRev.71.38

\bibitem{Yang:1947ud}
C.~N.~Yang,
Phys.\ Rev.\  {\bf 72} (1947) 874.
doi:10.1103/PhysRev.72.874

\bibitem{Heckman:2014xha}
J.~Heckman and H.~Verlinde,
Nucl.\ Phys.\ B {\bf 894} (2015) 58
 doi:10.1016/j.nuclphysb.2015.02.018
[arXiv:1401.1810 [hep-th]].

\bibitem{Buric:2015wta}
M.~Buri\'{c} and J.~Madore,
Eur.\ Phys.\ J.\ C {\bf 75} (2015) no.10,  502
doi:10.1140/epjc/s10052-015-3729-6
 [arXiv:1508.06058 [hep-th]].

\bibitem{Sperling:2017dts}
M.~Sperling and H.~C.~Steinacker,
J.\ Phys.\ A {\bf 50} (2017) no.37,  375202
doi:10.1088/1751-8121/aa8295
[arXiv:1704.02863 [hep-th]].


\bibitem{Buric:2017yes}
M.~Buri\'{c}, D.~Latas and L.~Nenadovi\'{c},
arXiv:1709.05158 [hep-th].

\bibitem{Steinacker:2016vgf}
  H.~C.~Steinacker,
  JHEP {\bf 1612} (2016) 156
  doi:10.1007/JHEP12(2016)156
  [arXiv:1606.00769 [hep-th]].

\bibitem{Hammou:2001cc}
A.~B.~Hammou, M.~Lagraa and M.~M.~Sheikh-Jabbari,
Phys.\ Rev.\ D {\bf 66} (2002) 025025
doi:10.1103/PhysRevD.66.025025
[hep-th/0110291].

\bibitem{Madore:1991bw}
  J.~Madore,
  Class.\ Quant.\ Grav.\  {\bf 9} (1992) 69.
  doi:10.1088/0264-9381/9/1/008

\bibitem{Hoppe} 
J. Hoppe,
Quantum theory of a relativistic surface,
Ph.D. Thesis, MIT, Advisor, J.
Goldstone (1982) 

\bibitem{Kovacik:2013yca}
S.~Kov\'{a}\v{c}ik and P.~Pre\v{s}najder,
J.\ Math.\ Phys.\  {\bf 54} (2013) 102103
doi:10.1063/1.4826355
[arXiv:1309.4592 [math-ph]].

\bibitem{Jurman:2013ota}
D.~Jurman and H.~Steinacker,
JHEP {\bf 1401} (2014) 100
doi:10.1007/JHEP01(2014)100
[arXiv:1309.1598 [hep-th]].

\bibitem{Szabo:2009tn}
  R.~J.~Szabo,
  Gen.\ Rel.\ Grav.\  {\bf 42} (2010) 1
  doi:10.1007/s10714-009-0897-4
  [arXiv:0906.2913 [hep-th]].

\bibitem{Castelino:1997rv}
  J.~Castelino, S.~Lee and W.~Taylor,
  Nucl.\ Phys.\ B {\bf 526} (1998) 334
  doi:10.1016/S0550-3213(98)00291-0
  [hep-th/9712105].

 \bibitem{Jurco:2000ja}
 B.~Jurco, S.~Schraml, P.~Schupp and J.~Wess,
 Eur.\ Phys.\ J.\ C {\bf 17} (2000) 521
 doi:10.1007/s100520000487
  [hep-th/0006246].
 
 \bibitem{Madore:2000aq}
   J.~Madore,
   ``An introduction to noncommutative differential geometry and its physical applications,''
   Lond.\ Math.\ Soc.\ Lect.\ Note Ser.\  {\bf 257} (2000) 1.

 \bibitem{Grosse:1993uq}
 H.~Grosse and P.~Presnajder,
 Lett.\ Math.\ Phys.\  {\bf 28} (1993) 239.
 doi:10.1007/BF00745155



\bibitem{Wallet:2016ilh}
  J.~C.~Wallet,
  Nucl.\ Phys.\ B {\bf 912}, 354 (2016)
  doi:10.1016/j.nuclphysb.2016.04.001
  [arXiv:1603.05045 [math-ph]].

\bibitem{Vitale:2012dz}
  P.~Vitale and J.~C.~Wallet,
  JHEP {\bf 1304} (2013) 115
   Addendum: [JHEP {\bf 1503} (2015) 115]
  doi:10.1007/JHEP04(2013)115, 10.1007/JHEP03(2015)115
  [arXiv:1212.5131 [hep-th]].

\bibitem{Vitale:2014hca}
  P.~Vitale,
  Fortsch.\ Phys.\  {\bf 62} (2014) 825
  doi:10.1002/prop.201400037
  [arXiv:1406.1372 [hep-th]].
  
  \bibitem{DeBellis:2010sy}
  J.~DeBellis, C.~Saemann and R.~J.~Szabo,
  JHEP {\bf 1104} (2011) 075
  doi:10.1007/JHEP04(2011)075
  [arXiv:1012.2236 [hep-th]].
  
  \bibitem{vilenkin} 
  N.~Ja.~Vilenkin, 
  ``Special Functions and the Theory of Group Representations,''
  Transl. Math. Monogr., vol. 22, AMS, 1968 
  
  \bibitem{VanProeyen:1999ni}
  A.~Van Proeyen,
  ``Tools for supersymmetry,''
  Ann.\ U.\ Craiova Phys.\  {\bf 9} (1999) no.I,  1
  [hep-th/9910030].
  
  \bibitem{Aschieri:2012vf}
  P.~Aschieri and L.~Castellani,
  Gen.\ Rel.\ Grav.\  {\bf 45} (2013) 581
  doi:10.1007/s10714-012-1488-3
  [arXiv:1205.1911 [hep-th]].

\bibitem{Gere:2013uaa}
  A.~G\'er\'e, P.~Vitale and J.~C.~Wallet,
  Phys.\ Rev.\ D {\bf 90} (2014) no.4,  045019
  doi:10.1103/PhysRevD.90.045019
  [arXiv:1312.6145 [hep-th]].

\bibitem{Alekseev:2000fd}
  A.~Y.~Alekseev, A.~Recknagel and V.~Schomerus,
  JHEP {\bf 0005} (2000) 010
  doi:10.1088/1126-6708/2000/05/010
  [hep-th/0003187].


 
 
  
  
 \end{thebibliography}
\end{document}